\documentclass[prl,twocolumn,floatfix]{revtex4-1} 
\usepackage{url}
\usepackage[pdfpagemode=UseNone,pdfstartview=FitH,colorlinks=true]{hyperref}

\begin{document}
\title{Does Collective Genetic Regulation exist?}

\author{J. M. Deutsch}
\email{josh@ucsc.edu}
\affiliation{Department of Physics, University of California, Santa Cruz CA 95064} 
%\institute{
%       Joshua M Deutsch \at
%       Department of Physics, University of California, Santa Cruz, CA 95064, USA
%       \email{josh@ucsc.edu}
%       }
\date{\today}

\begin{abstract}
Does regulation in the genome use collective behavior, similar to the way the brain
or deep neural networks operate? Here I make the case for why having a genomic network capable of a high
level of computation would be strongly selected for, and suggest how it might arise from
biochemical processes that succeed in regulating in a collective manner, very different than the usual
way we think about genetic regulation.
\end{abstract}
\maketitle
 
\section{Introduction}

The operating system for Linux (the Linux kernel) is over 27 million lines. If you randomly
duplicate a single line of it, recompiling it will likely result in a fatal error, so that it is no longer be a viable 
operating system. In contrast, if you duplicate the entire genome of many species of salamanders and
allow the polyploid egg to develop~\cite{schmid2015polyploidy}, it produces viable individuals that are even able to produce offspring
themselves. The same comparative lack of flexibility is found in most human-made technology, from cars to bicycles
to digital thermometers. And this distinguishes our technology from biological organisms that show
tremendous flexibility in alteration of their blueprints, yet still frequently
result in a viable organism.

Is this extreme flexibility something intrinsic to biology's basic design, or
is this a property that has required evolution over billions of years? I will argue
that the answer is the latter, and that this flexibility is crucial in allowing
life to have evolved so many diverse and sophisticated traits, including our level of intelligence.
This flexibility is closely related to the notion of genomic intelligence, the idea that
the apparatus controlling the regulation of genes is performing
sophisticated computations in order to evolve an organism efficiently.

This is different than genomic robustness~\cite{barkai1997robustness,savageau1985mathematics,savageau1985theory,masel2009robustness}, 
which is the idea that the genome can withstand
mutations and show little or no change in phenotype. It is changes in phenotype that
lead to adaptation and increased fitness in response to environmental change. Therefore it is important
to be able to produce mutations to adapt, without destroying viability in that process. That is to
say, such organisms are highly {\em evolvable}~\cite{pigliucci2008evolvability,kirschner1998evolvability}.

\section{The necessity for genomic intelligence}

The claim that genomic regulation is intelligent, might appear too vague to be a useful
way of guiding research. Let us delve further into how exactly intelligence is being
used in this biological context, and how it relates to ideas in artificial intelligence.

Intelligence can be defined as using knowledge to further goals. An example of neural
intelligence is to use the pattern of photons impinging on one's retina to determine that
a tiger is coming towards you, and that the best course of action is to run away. 
The way that we identify patterns is largely through learning. We are given many examples
of patterns and are often told of their classification, say, tigers, bicycles, and
cars. These are examples of ``supervised learning"~\cite{hertz1991introduction}. Let us try to determine if there is a
genomic analogy to this kind of learning.

Instead of inputs being visual patterns, the genome will have a large number of external inputs from
cell signals such as growth factors (receptor tyrosine kinases), ion channels, or adhesion
sensing~\cite{cooper2007cell}. The information from these signals are communicated to the nucleus. This changes
its state, causing it to alter the outputs of this process, that is the proteins being produced. 
This can tell the
cell, for example, to stop growing or to form a synapse with another cell. There is a lot of regulatory
machinery in the nucleus that is computing these outputs.

This has a fairly close analogy to the way the brain, or an artificial intelligence (AI), learns patterns
such as described above~\cite{hertz1991introduction}.
The difference is that there is no direct supervision going on. The inputs are given, but it is not
obvious how the cell is supposed to know the expected outputs.

In principle, evolution will eventually be able to adjust regulation so that outputs correctly match the inputs.
Mutations of the genome will alter the proteins that are produced in response to a given input.
These can cause developmental changes and thus changes in the developed organism, affecting
its fitness. This will cause an evolution towards fitter cell outputs. But this is a much more indirect
and therefore
inefficient process than what can be achieved with supervised learning. Hence evolution through
genetic change would appear to proceed through a much less efficient algorithm than what is achievable
through supervised learning algorithms such as are employed in AI.

From experience with human made designs, the most difficult hurdle to overcome is the lack of robustness. If 99\% of the
time a mutation is lethal, it makes evolution inefficient. If instead, mutations cause tiny effects, then
although an organism might be viable, it will be functionally almost identical to its progenitor.
This would be a problem associated with inflexibility.
What is needed is intelligent genomic regulation, where even with large changes to inputs,
the outputs will be different but not in ways that  cause the organism to fail at development, thus
allowing the potential for many more beneficial mutations.

When the genome mutates, new patterns of signals impinge on the cell
(as well as the possibility of new kinds of protein signals). The genetic network
must be able to generalize well so as to respond efficaciously to these new signals.
And not only must the genome respond to mutation, but to different gene alleles that
occur with sexual reproduction. Intelligent responses would also allows for much improved regenerative
capacity~\cite{brockes2001regeneration}.

But this kind of intelligent regulation as described,  is not an example of
supervised learning, because there is no cell training where desired cell outputs are given. 
Instead the cell likely uses an empirical approach, using signals impinging on it, and
learning how actions that it takes affect the environment around it. This is
quite similar to a well known technique in AI, reinforcement learning~\cite{watkins1992q}. 

%Feedback is necessary to respond to new signal patterns caused by mutations,
%because of the extreme complexity of biochemical networks, which preclude predicting the effects
%of all interactions. Genomic regulation is an example of a complicated dynamical system and doing such prediction
%would, in effect, require the cell simulating itself~\cite{beckage2013more}. Instead in order to compute
%how to adjust outputs, the cell likely uses an empirical approach, using signals impinging on it,
%and learning how actions that it takes affect the environment around it. This is
%quite similar to a well known technique in AI, reinforcement learning~\cite{watkins1992q}. 

Regulatory feedback goes far beyond the nucleus including signals coming from
other cells. If a mutation has had a positive effect in some respects but has caused 
a side-effect, for example a decrease in cell adhesion, this can be
signaled back to the nucleus and the genetic network can compute what measures are necessary to
ensure cells are correctly bound together. What is important from the standpoint of evolutionary
efficiency is allowing certain changes to happen without destroying the viability of
an organism. For example, mutations leading to greater intelligence are difficult
for a number of reasons. Blood supply must be increased, the skull must also expand,
but more importantly, there is an incredible complexity involved in
neural processing, and small changes would be lethal for most architectures (such as the Linux kernel)
without very sophisticated rules for how development should proceed. If all of the
components needed to co-evolve require separate mutations, this would tremendously stifle
the possibility of useful genetic changes. Instead there is likely a great deal
of intelligence applied in the way say, angiogenesis or neural growth, progresses when tissue receives signals that
require it. 

%This view of development suggests that the speed of evolution is limited by the computational capacity of the genome. 
The above argument concerns genomic intelligence to allow for biological flexibility.
But a smarter genome will speed up  evolution in other important ways.
Intelligent agents use the past to better predict optimal future actions to take. The genome
does not have a direct record of its ancestor's history. It does not know directly if there
had been a prior ice age that appears to be re-emerging again. But it does contain a great
deal of spare capacity to store away information from previous generations, that can then
be used in development~\cite{goldschmidt1933some}. For example, if there is a change to the environment,
say a drop in average temperature, a mutation can switch on a gene that grows hair.
The process of growing hair does not need to re-evolve from scratch. This example has no
concept of directionality or ordering in time, just that there are useful genes that can be turned on, if needed.
However accessing the timing of previous genetic changes gives the genome the opportunity to
better predict and optimize developmental response. The study of heterochrony\cite{klingenberg1998heterochrony},
how timing in development is influenced by evolution, shows that the genome does indeed
have access to some, admittedly crude, approximation to the order in which genetic change happened.

Therefore the genome can utilize information about its present and past states, to make
decisions on how to develop. For example, a gene to suppress the production of hair
may be able to access the fact that its recent ancestors appear to have a tendency of becoming
less hirsute. This would suggest to the genomic network that a mutation to suppress hirsuteness have
its effects enhanced, creating
an acceleration in evolution of a hairless phenotype. Although directed mutations are not
normally possible from a physical perspective, processing of past information contained in the genome can lead
to similar evolutionary behavior.

The above discussion illustrates that a highly intelligent genome, capable of
performing complex prediction and inference, allows an organism to respond
to change more efficiently than one's with lesser computational sophistication.
Evolution of such complex genetic regulatory machinery
will then make an organism more highly evolvable~\cite{wagner1996perspective,pigliucci2008evolvability}. It would
appear likely that such machinery has been selected for, and is why biology
contains the tremendous flexibility in development discussed above, in
comparison to human technology. Despite the vast amount that we have learned from
evolutionary developmental biology, this view does not appear to be widespread.
Evolvability has been studied
extensively~\cite{kirschner1998evolvability,pigliucci2008evolvability},
and a large number of traits associated with it have been studied and described.
For example, weak linkage~\cite{conrad1990geometry} (e.g. between different cellular processes),
exploratory mechanisms (e.g. adaptive immunity), and genomic compartmentation~\cite{wagner1996perspective}.
All of these traits are logical guidelines you would also expect in a complex human designed
machine, or software, that was built to be upgradable. But these guidelines are not nearly enough
to actual build the algorithms necessary for the machine to function. The necessary feature
for evolvability is intelligence.

This then begs the question, of how the genome would be able to perform this level
of sophisticated computation, which I turn to now.

\section{Mechanisms for computation in the genome}

Only about 3\% of the genome is translated into proteins. Yet until recently, it was
only these portions of the genome that were considered as having an important biological
function. A great deal of recent effort has been devoted recently to studying
the function of the other 97\%, and it has been shown to have a very large number
of functions \cite{snyder2020perspectives,davis2018encyclopedia}. However given the amount of DNA, and
the complexity of its functions, it is still far from being well understood.
Because this DNA is not directly producing proteins, its function will be
to regulate the gene-coding portions of the genome. This gives the genome
a much larger amount of information to utilize in computation and
potentially lead to much more complicated regulatory mechanisms. This
brings us now to the heart of this discussion: is gene regulation related to AI?

We now are living at the beginning of a new age in computer science, where software and hardware
utilizing ``Deep Learning", or artificial neural networks~\cite{deng2014deep,lecun2015deep} has outstripped older
methods in machine learning. Could similar ideas be important in the way that
the genome performs regulation?

The distinguishing feature of this kind of architecture, is
that computation is done collectively, with many inputs impinging on a single element, like that of
a neuron. This idea dates back to 1958, which is when the ``perceptron" was invented~\cite{rosenblatt1958perceptron}.
and gave rise to work in connectist models, which essentially stack perceptrons to create
more powerful learning algorithms~\cite{mcclelland1986parallel}. These models have now
solved problems that had hitherto been considered intractable, such as speaker independent
speech recognition~\cite{deng2013new}.

This biologically inspired architecture is much more general and simpler than the
earlier and more traditional rule based approaches. Instead of attempting to hand-code 
the architecture of a task, such as translating French into English, English sounds
are given as inputs, and French as outputs~\cite{deng2013new}. The strength of the connections are
algorithmically adjusted until the network has learned the task.

The distinction between the traditional rule based algorithms, and deep learning, is the
idea of collective computation. A single connection between two neurons, is
serving many simultaneous functions. Its purpose is manifold. It is only when
combined holistically with the other neurons and connections, that a precise computation emerges.
Single connections can be severed, and this normally has little effect
on performance. In this way, neural network models are robust. In contrast,
a traditional digital circuit, or the source code for the Linux kernel, is
extremely fragile.

Genetic regulation is traditionally thought of as a network of regulatory elements
that can for example, enhance or silence, transcription of their associated genes.
It is often considered to be boolean and behaves similarly to traditional
digital circuits, for example AND and NOT gates. There is little doubt
that this kind of regulation plays an important role in gene regulation and evolution~\cite{wray2007evolutionary},
however, it is the purpose of this section to point out that the other 97\% of
of our genome, does not fit neatly into this traditional picture, and can potentially be used for
{\em collective regulation} which would share many features in common with
collective computation, like Deep Learning. It has the potential for greatly
increasing the intelligence of the genome, and therefore would be evolutionarily
selected for. Analogies with connectist models have been proposed
several decades ago\cite{mjolsness1991connectionist}.

Non-coding RNA (ncRNA) is expressed with an abundance of about one tenth that of mRNA,
but this depends strongly on cell type~\cite{cabili2011integrative}. 
Physical arguments~\cite{deutsch2018computational} give a collision time between different ncRNA of approximation $0.25 s$. But the
half-life a ncRNA in the nucleus is of the order of 30 minutes~\cite{lee2012epigenetic}. Therefore
there is plenty of time for ncRNA to interact before being degraded.

It is possible~\cite{deutsch2018computational} to come up with a theoretical analysis of how interactions between the RNA molecules,
and a mechanism for their creation and degradation, can map onto a model of collective computation,
such as a Boltzmann machine or Hopfield model~\cite{BoltzmannMachine,hopfield1982neural}. The
features of this model are as follows
\begin{itemize}
   \item[a] The equilibration of $N$ different RNA species that undergo pairwise binding and unbinding to each other with a set of
      equilibrium constants. At one time, some fraction of every species will be bound to another RNA molecule.
   \item[b] A creation rate (due to RNA polymerase II) for an RNA species that depends on the fraction of bound to unbound RNA
      for that species.
   \item[c] Degradation of RNA on a much longer timescale than than the interactions between the RNA
      molecules.
\end{itemize}
The interaction strengths between any two RNA molecule are weak and they bind promiscuously to each
other. By adjusting the equilibrium constant for binding, the system can evolve to have
an arbitrary relationship between input signals and the outputs that are produced. The inputs to
cells through signalling will affect concentrations of RNA molecules in the nucleus, this is then
processed by the above mechanism, to produce output mRNA molecules that will then be transcribed to
proteins. Item b above specifies a creation rate as a function of bound to unbound RNA. This
function is required to have a particular kind of sigmoidal shape in order for this model to
map onto a neural network model. There is evidence that this kind of creation is sometimes
utilized by noncoding RNA~\cite{sigova2015transcription,takemata2016local,takemata2017role}.

The basic framework for regulation of this kind, is that there is a lot of weak 
binding and unbinding between different biomolecules in a cell's environment. The sum total of these interactions 
would seem to serve no useful purpose. However coupling this with a creation rate that depends
on the bound fraction of such molecules, can in principle, perform sophisticated collective computation.

This particular model is unlikely to be precisely what is found in the cell nucleus. However
it points out that there are mathematically viable mechanisms based on the known molecular
biology of the cell that can in principle perform sophisticated gene regulation.

One of the arguments that has been used to dismiss the 97\% of DNA that is non-coding, is that
it is often not evolutionarily conserved. 
In comparison with traditional regulation, this collective mechanism is quite robust to mutation~\cite{deutsch2018computational}. 
Therefore one would expect that
to optimize evolution, a much higher mutation rate is desirable as is often seen for non-coding
RNA~\cite{ENCODEpilot}.

\section{Discussion}

Unfortunately at the moment, there is no strong evidence for collective regulation. There is evidence for some of
the pieces, such as extensive RNA-RNA interactions in humans~\cite{sharma2016global}, that can
detect the formation of inter-RNA duplexes, stronger interactions than would be optimal for
the weak interactions described above. But the general mechanism described to achieve
collective regulation could in principle be accomplished by a large diversity of different
kinds of molecules, including proteins. Weak interactions combined with control over
creation rates are the main two requirements. 

Genome wide association studies
%(GWAS) 
are widely used to understand the
genetics of a large variety of diseases. It has already become clear that single genes
cannot predict risk, and instead this is controlled by a large number of different regions,
90\% of these are  non-coding regions~\cite{diaz2020omics,gallagher2018post}.
This is circumstantial, but hardly compelling evidence for regulation being
collective. 

Biologist are very good at finding specific biochemical interactions and have
uncovered an enormous amount about regulation of the genome. However it would be much harder to make sense
of thousands of interactions simultaneously to try to uncover a new deep-learning-like mechanism for
genomic regulation. 
However it is argued here that this sort of architecture is plausible biochemically and would
be selected for evolutionarily. Therefore it deserves some effort to try to devise
experimental methods to test for it.
If collective regulation turns out to be key, this will have
significant implications for the future direction of a lot of research.
This would have scientific
benefit not only in biology, but inform us on how to design more flexible software and hardware.

\section{Acknowledgements}

The author thanks William Sullivan for useful discussions.
This work was supported by the Foundational Questions Institute \url{<http://fqxi.org>}.

%\clearpage
\bibliographystyle{apalike}      % mathematics and physical sciences
\bibliography{genomic_intelligence}

\end{document}